\documentclass[lettersize,journal]{IEEEtran}

\makeatletter
\renewcommand\@biblabel[1]{$[{#1}]$}

\usepackage{lipsum}
\def\cleardoublepage{\clearpage\if@twoside \ifodd\c@page\else
	\thispagestyle{empty}%
	\hbox{}\newpage\if@twocolumn\hbox{}\newpage\fi\fi\fi}
\makeatother

\usepackage {booktabs}
\usepackage {layouts}
\usepackage {multirow}
\usepackage {threeparttable}
\usepackage [noend]{algorithmic}
\usepackage[ruled,linesnumbered,vlined]{algorithm2e}

\usepackage{caption}
\usepackage{amsmath}
\allowdisplaybreaks 
\usepackage{amssymb}
\usepackage{graphicx}
\usepackage{hyperref}
\usepackage{url}
\usepackage{xcolor}

\usepackage{subfigure}
\usepackage{cite}
\DeclareCaptionLabelFormat{mylabel}{\MakeUppercase{#1}~#2}

\newcommand{\topcaption}{%
	\setlength{\abovecaptionskip}{0cm}%
	\setlength{\belowcaptionskip}{0cm}%
	\caption}

\begin{document}

\title{ICPS: Real-Time Resource Configuration for Cloud Serverless Functions Considering Affinity}

\author{Long~Chen,
        Xinshuai~Hua,
        Jinquan~Zhang, 
        Wenshuai~Li,
        Xiaoping~Li,~\IEEEmembership{Senior Member,~IEEE,}
        and Shijie~Guo
        \IEEEcompsocitemizethanks{
        \IEEEcompsocthanksitem This work is supported by the National Key Research and Development Program of China (No. 2022YFB3305500), the National Natural Science Foundation of China (No. 62273089, 62102080), the Fundamental Research Funds for the Central Universities (No.2242022R10017).
        \IEEEcompsocthanksitem Long Chen, Xinshuai Hua, and Wenshuai Li are with the School of Computer Science and Engineering, Southeast University, Nanjing, 211189, China, 
        and also with the Key Laboratory of New Generation Artificial Intelligence Technology and Its Interdisciplinary Applications, Southeast University, Ministry of Education, Nanjing 211189, China.
        Email: chen\_long@seu.edu.cn.
        \IEEEcompsocthanksitem Jinquan Zhang (Corresponding author) and Xiaoping Li are with the School of Computer Science, Guangdong University of Technology, Guangzhou, 510006, China.
        E-mail: zjq8860@gdut.edu.cn.
        \IEEEcompsocthanksitem Shijie Guo is with the College of Electronic Engineering, National University of Defense Technology, Hefei, 230037, China. E-mail: gsj53@sina.cn.
        }
}

\markboth{Journal of \LaTeX\ Class Files,~Vol.~14, No.~8, August~2021}%
{Shell \MakeLowercase{\textit{et al.}}: A Sample Article Using IEEEtran.cls for IEEE Journals}


\maketitle

\begin{abstract}
Serverless computing, with its operational simplicity and on-demand scalability, has become a preferred paradigm for deploying workflow applications.
However, resource allocation for workflows, particularly those with branching structures, is complicated by cold starts and network delays between dependent functions, significantly degrading execution efficiency and response times.
In this paper, we propose the Invocation Concurrency Prediction-Based Scaling (ICPS) algorithm to address these challenges.
ICPS employs Long Short-Term Memory (LSTM) networks to predict function concurrency, dynamically pre-warming function instances, and an affinity-based deployment strategy to co-locate dependent functions on the same worker node, minimizing network latency.
The experimental results demonstrate that ICPS consistently outperforms existing approaches in diverse scenarios. The results confirm ICPS as a robust and scalable solution for optimizing serverless workflow execution.
\end{abstract}


\section{Introduction}

\IEEEPARstart{A}{pplications} in the cloud commonly consist of several components \cite{li2022serverless} (e.g., real-time data processing and web backend services).
These components form a workflow, which means that they are orchestrated sequentially or in parallel \cite{8756833}, and can be executed according to predefined steps once a request is received.
However, the number of requests from users fluctuates over time.
Thus, it is essential to dynamically configure resources for various components.
Serverless computing has emerged as an ideal choice for executing these applications due to its simplicity and on-demand scalability \cite{9190031}, which can effectively reduce resource waste and user costs.

In serverless computing, each component in an application is called a function.
Each invoked function is executed in a container, also called a function instance.
However, if a function is not invoked for a period of time, the container may be released to reduce resource usage.
When the function is invoked again, a new container is launched. This causes a delay known as cold start \cite{golec2023cold}, which can significantly impact the function execution time.
Allocating substantial resources to the function can effectively reduce cold starts while leading to higher resource usage.
The cold start has a more significant impact on workflows \cite{burckhardt2022netherite}.
Each function in a workflow faces cold start issues and may be invoked by other functions.
Therefore, the delay caused by these functions accumulates in the workflow \cite{erbati2021application}, significantly increasing the response time of the workflow \cite{Xanadu}.
Consequently, configuring resources for functions to address the cold start issue is essential to reduce the response time of the workflow and resource usage.

In this paper, the response time minimization problem for workflows with minimal resource consumption is studied.
This problem has three main challenges:
(1) A workflow application may include branches due to its business logic, resulting in multiple sub-graphs and unbalanced resource requirements across functions.
It is a challenge to allocate resources in a fine-grained manner for different functions.
(2) The concurrency of different types of requests (i.e., sub-graphs) varies over time.
This makes it difficult to accurately predict resource requirements and dynamically adjust resource allocation for functions in real time.
(3) Function instances are deployed on worker nodes, such as virtual machines.
Network latency occurs when functions deployed on different worker nodes need to communicate.
Minimizing network latency poses a challenge in deploying functions and allocating requests efficiently.

Previous studies tackling the cold start problem concentrate on individual functions\cite{Implications}, overlook the invocation relationships among functions, and consequently restrict their potential to enhance workflow response times.
To address the unique challenges in workflow applications, we propose the Invocation Concurrency Prediction-Based Scaling Algorithm (ICPS) in this paper to configure resources for functions in real time.
Our contributions are as follows.
\begin{itemize}
    \item We examine the problem of minimizing response time for workflows with branching structures, considering the network latency between function instances deployed on different worker nodes to enhance the practicality of our study.
    \item We adopt a Long Short-Term Memory (LSTM) network to predict the concurrency of different sub-graphs, based on which the concurrency of functions is dynamically determined to pre-warm function instances to avoid frequent cold starts and excessive resource waste.
    \item We design an affinity-based instance deployment method, which reduces network latency by co-locating instances of functions with dependency relationships on the same worker node.
\end{itemize}

This paper is structured as follows.
Section 2 reviews the related work.
Section 3 presents a detailed description and formulation of the studied problem.
The proposed algorithms are detailed in Section 4.
The experimental results are given in Section 5,
followed by the conclusions and future research in Section 6.

\section{Related Works}
The cold start of function instances is a critical issue in serverless computing and has attracted considerable attention.
Various studies have been conducted to address the cold start problem.

Some studies focus on reducing the occurrence of cold starts.
In serverless computing, invoking a function may trigger the creation of a new function instance, leading to a cold start \cite{Implications}. 
Sahar et al. \cite{MOAKHAR2024103890} proposed the EF-TTC algorithm to optimize function scheduling and placement in edge computing environments.
Li et al. \cite{Performance-First} developed a deployment model to balance performance and resource efficiency for serverless workflows.
Solaiman et al. \cite{solaiman2020wlec} analyzed the causes of cold starts on the OpenLambda platform and reviewed techniques to minimize container initialization times.
Shen et al. \cite{shen2021defuse} introduced Defuse, a dependency-based function scheduler designed to mitigate cold starts on Function-as-a-Service (FaaS) platforms.
Xu et al. \cite{8975850} utilized LSTM networks \cite{llllstm} to predict request trigger times, enabling dynamic pre-warming of function instances.
Gunasekaran et al. \cite{gunasekaran2020fifer} proposed an adaptive resource management framework to efficiently manage function chains on serverless platforms.
Kumari et al. \cite{kumari2024acpm} proposed ACPM, an integrated model that dynamically pre-configures containers at runtime to reduce cold starts.
Roy et al. \cite{roy2022icebreaker} proposed a technique to alleviate the costly overhead during the pre-warming of the function instance.

Other studies focus on optimizing the startup process of function instances to minimize initialization delays.
Abad et al. \cite{Package} introduced a package-aware scheduling algorithm that accelerates instance startup by prioritizing function nodes with significant dependency relationships.
Mahajan et al. \cite{Exploiting} optimized the instance startup process by precreating networks for function containers, reducing the time required for network setup.
Du et al. \cite{du2020catalyzer} introduced a function instance snapshot and recovery technique based on gVisor \cite{wang2022performance}. This technique allows for the rapid creation of function instances by either fully restoring snapshots or sharing memory data with running function instances.

In conclusion, existing studies overlook the invocation dependencies of functions in workflows with branching structures.
Furthermore, they rarely consider the co-location of function instances to minimize network latency.
In this paper, we address these limitations by considering workflow branches and instance co-location, aiming to minimize workflow response time and reduce resource waste.

\section{Problem Description}

\subsection{Assumptions}
Considering the complexity of the studied problem, some assumptions are made before formulating the problem. The details are as follows:
\begin{enumerate}[]
    \item The execution time and resource requirements of each function are fixed and known in advance.
    \item Function instances running on worker nodes (virtual machines) operate in a non-preemptive manner. This means that they cannot be replaced by other functions until they are released.
    \item Each instance handles only one function invocation at a time, ensuring exclusive execution.
    \item The network delay between worker nodes is constant, regardless of data size, node location, or network load.
\end{enumerate}

\subsection{Problem Formulation}
\begin{figure}[htbp]
    \centering
    \includegraphics[scale = 0.75]{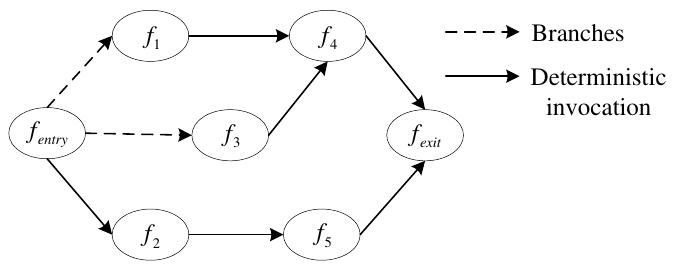}
    \topcaption{An example of workflow application}
    \label{Fig:workflowapplication}
\end{figure}

A workflow application is a directed acyclic graph (DAG) $G = (V, E)$.
$V = \{f_{entry}, f_1, f_2, \ldots, f_{N}, f_{exit}\}$ is the set of functions,
where $N$ is the number of functions in $G$ except for the unique entry function $f_{entry}$ and the unique exit function $f_{exit}$.
$E$ is the dependency relationship among these functions.
Each request the user submits is a workflow, which is a sub-graph of $G$ due to the branches in the workflow application, as shown in Figure \ref{Fig:workflowapplication}.
Suppose that $WF = \{wf_1, wf_2, \ldots, wf_M\}$ is all the workflows submitted in duration $D$, and the subscripts are arranged in a non-decreasing order of arrival time $T^{arr}_i$.
Workflow $wf_i = \{V_i, E_i\}$, where $V_i=\{f^i_{entry}, f^i_1, f^i_2,...,f^i_{n_i}, f^i_{exit}\}$ is the set of functions invoked in workflow $wf_i$.
Since there is only one exit function, the finish time of function $f^i_{exit}$ is also the finish time of $wf_i$, denoted as $T^{end}_i$.
Therefore, the response time $T^{resp}_i$ of $wf_i$ is the gap between $TW^{arr}_i$ and $TW^{end}_i$, i.e.,
\begin{gather}
    T^{resp}_i = T^{end}_i-T^{arr}_i.
\end{gather}
The response time $T^{resp}_i$ includes the total execution time $T^{exec}_i$ of all functions in the critical path, the waiting time for execution (caused by queuing or cold starts), and the network latency caused by data transmission.
If the invocation of a function is allocated to an existing instance, the cold start time is 0 due to instance reusing.
In this paper, the average response efficiency $\phi_{resp}$ is used to evaluate the performance of executing workflows, calculated by Formula \eqref{eq:resp_uti}.
\begin{gather}
    \phi_{resp} = \frac {\sum_{i=1}^M T^{exec}_i} {\sum_{i=1}^M T^{resp}_i} \label{eq:resp_uti}
\end{gather}

Functions are executed in containers, and the containers are deployed on worker nodes (i.e., virtual machines).
The number of worker nodes created in duration $D$ is $W$.
$WN = \{wn_1, wn_2, \ldots, wn_W\}$ is the set of worker nodes.
The memory of the $w^{th}$ worker node $wn_w$ is $wn^{mem}_w$.
Suppose $wn^{mem}_{w,t}$ is the memory occupied by containers in the worker node $wn_w$ at time $t$,
Formula \eqref{eq:memory_limitation} should be satisfied since the resource in the worker node $wn_w$ is limited.
\begin{gather}
    wn^{mem}_w \geq wn^{mem}_{w,t} \label{eq:memory_limitation}
\end{gather}
If two function instances are co-located on the same worker node, the data transmission time is negligible. Otherwise, a fixed network latency of $d$ is incurred.

Suppose that $P$ is the number of all created function instances of duration $D$ and $I_k$ is the $k^{th}$ instance.
The resource utilization $\phi_{resource}$ is calculated by Formula \eqref{eq:resource_util}.
\begin{gather}
    \phi_{resource}= \frac {\sum_{k=1}^P C_k^{exec}} {\sum_{k=1}^P C_k^{total}} \label{eq:resource_util}
\end{gather}
where $C_k^{total}$ is the total resource consumption of the $k^{th}$ instance $I_k$, and $C_k^{exec}$ is the resource consumption of instance $I_k$ in the running time and cold start time.
Suppose that $TC_k^{total}$ is the life cycle time of instance $I_k$ (running time, cold start time, and idle time caused by reusing of instance $I_k$ are included), and $TC_k^{idle}$ is the idle time, $C_k^{total}$ and $C_k^{exec}$ can be calculated by Formulas \eqref{eq:total_deployed} and \eqref{eq:deployed}, respectively.
\begin{gather}
    C_k^{total} = TC_k^{total} \times C_k^{mem}\label{eq:total_deployed}\\
    C_k^{exec} = (TC_k^{total}-TC_k^{idle}) \times C_k^{mem} \label{eq:deployed}
\end{gather}
where $C_k^{mem}$ is the memory size of instance $k$, and $TC_k^{total}$ is the difference between the release time $TC_k^{kill}$ and $TC_k^{create}$ is the creation time of instance $I_k$, i.e.,
\begin{gather}
    TC_k^{total} = TC_k^{kill} - TC_k^{create}
\end{gather}

This paper aims to promote $\phi_{resp}$ and $\phi_{resource}$. Therefore, the objective of this paper is to minimize their product, as shown in Formula \eqref{eq:targgget}
\begin{gather}
    \min~\eta=\phi_{resp}\times\phi_{resource} \label{eq:targgget}
\end{gather}
The decision variables include the numbers of pre-warmed containers at each time interval $TI$,
the deployment locations of function instances, and the function instance chosen for executing each function invocation.

\section{Proposed Algorithms}
In this section, we first introduce the adopted system architecture, followed by a detailed description of the proposed Invocation Concurrency Prediction-Based Scaling Algorithm (ICPS).

\subsection{System Architecture}
\begin{figure}[htbp]
    \centering
    \includegraphics[scale = 0.6]{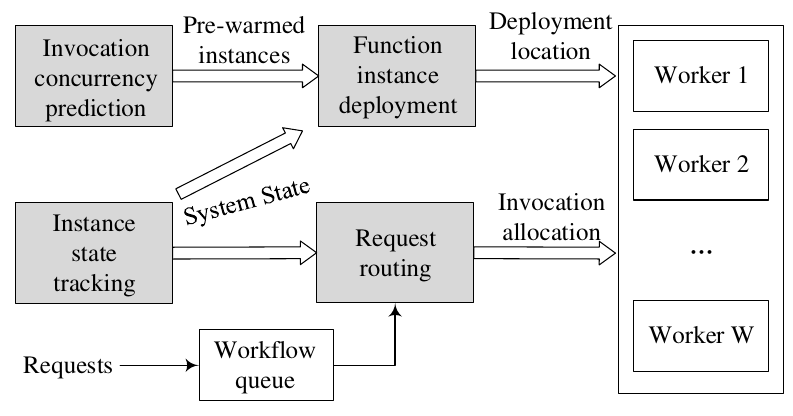}
    \topcaption{System architecture}
    \label{Fig:thirdalar}
\end{figure}

The adopted system architecture is shown in Figure \ref{Fig:thirdalar}.
The system architecture comprises four distinct modules, namely invocation concurrency prediction, instance state tracking, function instance deployment, and request routing.
\begin{itemize}
  \item Invocation concurrency prediction: The invocation concurrency prediction module is in charge of predicting the concurrency of each function. The result is adopted to determine the number of pre-warmed containers for the next time interval.
  \item Instance state tracking: This module collects status information for all function instances. The status information is used to select deployment locations in the function instance deployment module and to allocate invocations in the request routing module.
  \item Function instance deployment: This module is responsible for deploying newly created function instances, triggered by pre-warming or cold starts, onto worker nodes based on information provided by the instance state tracking module.
  \item Request routing: The request routing module aims to assign a suitable instance for function invocations based on the information provided by the instance state tracking module, thereby avoiding cold starts and network latency.
\end{itemize}

\begin{algorithm}[htbp]
    \begin{small}	
        \topcaption{Invocation Concurrency Prediction-based Scaling Algorithm (ICPS)}
        \label{alg:ICPS_frame}
        \KwIn{$H$: Historical data of workflow and function invocation, $Re$: Requests received in the next time interval}
        \KwOut{$\phi_{resp}$: Average response efficiency, $\phi_{resource}$: Resource utilization}
        \Begin{
            \ForEach{time interval}
            {
                $P \gets$ InvocationConcurrencyPrediction($H$)\;
                $S \gets$ The current system state\;
                FunctionInstanceDeployment($P$, $S$)\;
                Initialize queue $Q \gets Re$\;
                \While{$Q \neq \emptyset$}
                {
                    $F \gets$ The first invoked function in $Q$\;
                    $S \gets$ The current system state\;
                    RequestRouting(F, S)\;
                }
                Update $H$, $Re$\;
            }
            Compute $\phi_{resp}, \phi_{resource}$\;
            \Return $\phi_{resp}, \phi_{resource}$.
        }
    \end{small}
\end{algorithm}

The ICPS algorithm framework is shown in Algorithm \ref{alg:ICPS_frame}.
The input of ICPS is the historical data of the workflow and invocation of the function $H$, and the requests received in the next time interval $Re$.
The output includes the average response efficiency $\phi_{resp}$ and the resource utilization $\phi_{resource}$.
Firstly, the types and numbers of function instances need to be pre-warmed are generated based on the historical data at the beginning of each time interval to reduce the occurrence of cold starts.
Secondly, these functions are deployed on worker nodes based on the current state of the system to reduce the network latency caused by data transmission.
Subsequently, once a workflow (request) arrives, it is added to a queue.
The workflow is not removed from the queue until all its invoked functions are executed.
Each function invoked in workflows in the queue is routed to a function instance based on the state of the system, which may be an existing or a newly created one.

\subsection{Invocation concurrency prediction}

The purpose of invocation concurrency prediction is to determine the types and numbers of pre-warmed function instances for the next time interval.
The occurrence of cold starts can be significantly reduced if the number of pre-warmed instances exceeds the concurrency of functions in the next time interval.
However, this strategy leads to low resource utilization.
In contrast, if the number of pre-warmed function instances is fewer than the concurrency of function invocations, frequent cold starts increase the response time of workflows.
Therefore, it is critical to accurately predict the function invocation concurrency for determining the numbers of pre-warming function instances.

There is a close relationship between the concurrency of workflows and the concurrency of functions.
In an application, there are usually several types of workflows, denoted as $K_1, K_2, \dots, K_S$.
Within a time interval $t$, their concurrency levels are $Con^t_1, Con^t_2, \dots, Con^t_S$, respectively.
The concurrency of the application $Con^t$ is the sum of all kinds of workflow concurrency.
\begin{gather}
    Con^t = \sum_{s = 1}^{S} Con^t_s \label{eq:application_concurrency}
\end{gather}
In addition, the concurrency $Con^t(f_n)$ of function $f_n$ is the sum of the concurrency of workflows that invoke $f_n$, that is,
\begin{gather}
    Con^t(f_n) = \sum_{s = 1}^{S} Con^t_s \times R(f_n, K_s) \label{eq:function_concurrency} \\
    R(f_n, K_s) = \begin{cases}
        1, & f_n \in K_s \\
        0, & otherwise
    \end{cases}
\end{gather}

\begin{figure}[htbp]
    \centering
    \includegraphics[scale = 0.5]{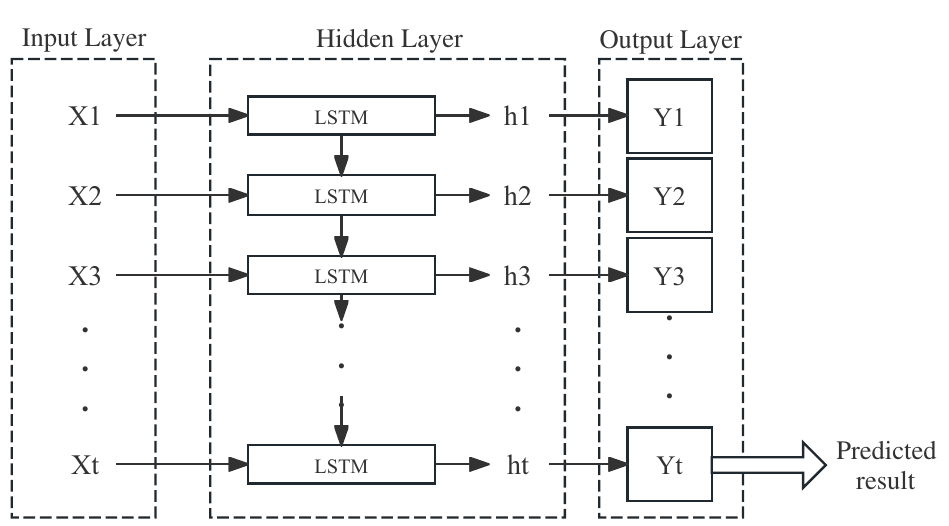}
    \topcaption{Structure of LSTM network}
    \label{Fig:LSTMMODEL}
\end{figure}

In this section, the Long Short-Term Memory (LSTM) network is adopted to predict the concurrency of workflows and functions,
the structure of which is shown in Figure \ref{Fig:LSTMMODEL}, including the input layer, hidden layer and output layer.
It receives vectors $(X_1, X_2, \ldots, X_t)$ as input and generates the final output $Y_t$ as the prediction result.
Specifically, three function invocation concurrency prediction strategies are proposed in this section.

\subsubsection{Full Path Concurrency Generation Strategy (FPCG)}

\begin{figure}[htbp]
    \centering
    \includegraphics[scale = 0.75]{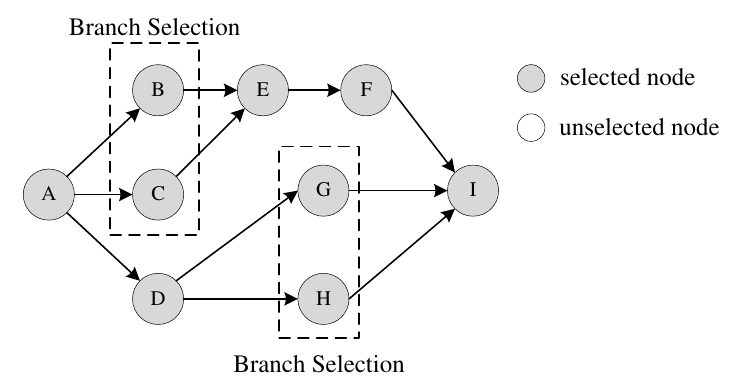}
    \topcaption{FPCG strategy}
    \label{Fig:FPCGgenerate}
\end{figure}

This strategy first predicts the workflow concurrency for the next time interval based on historical concurrency data.
Each data point $X_i$ in the input vector includes workflow types and their concurrency.
Although there are branches in workflows, functions in all invocation paths are pre-warmed to avoid cold starts, as shown in Figure \ref{Fig:FPCGgenerate},
regardless of the actual execution path taken by the workflow.
Therefore, the concurrency for each function is calculated using Formula \eqref{eq:application_concurrency}.

\subsubsection{Branch Path Concurrency Generation Strategy (BPCG)}
\begin{figure}[htbp]
    \centering
    \includegraphics[scale = 0.75]{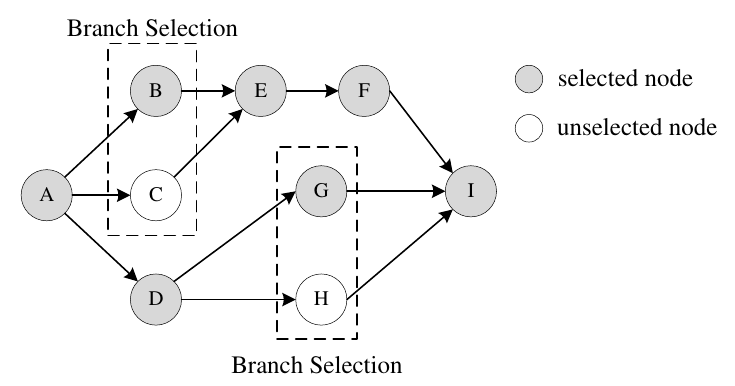}
    \topcaption{BPCG strategy}
    \label{Fig:BPCGgenerate}
\end{figure}

The BPCG strategy predicts the concurrency of the workflow in the same way as the FPCG strategy.
However, the concurrency of functions is calculated in a different way.
For workflows with branches, the BPCG strategy does not pre-warm all functions in the application according to the sum of workflow concurrency.
In contrast, it just pre-warms instances for actually invoked functions.
As in the subgraph shown in Figure \ref{Fig:BPCGgenerate}, function B is pre-warmed while function C is not.
The concurrency of functions is calculated using the Formula \eqref{eq:function_concurrency}.

\subsubsection{Concurrency History Statistics based Concurrency Generation Strategy (CHSCG)}

The CHSCG strategy performs statistical analysis on the frequency of different types of functions according to historical records, predicting the probability that each type of function is invoked in the next time interval, denoted by $q_1, q_2, \ldots, q_N$.
Suppose that the total number of function instances created in the last time interval is $TN$.
The numbers of pre-warmed instances are calculated by $q_n \times TN, \forall n=1,2, \ldots, N$.

\subsection{Instance state tracking}
\begin{figure}[!h]
    \centering
    \includegraphics[scale = 0.5]{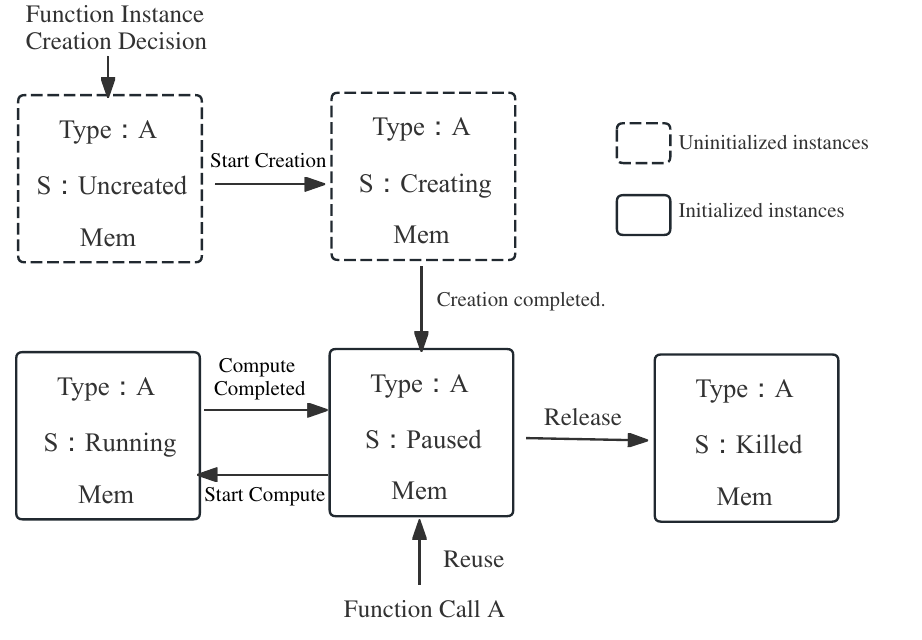}
    \topcaption{State transition of function instance}
    \label{Fig:status_change}
\end{figure}

The purpose of the function instance tracking is to monitor the state of all worker nodes and function instances in the system in real time,
which provides a basis for decision making for instance deployment and request routing.
The state of a worker node primarily includes its memory capacity and resource utilization.
However, function instances have multiple states, as shown in Figure \ref{Fig:status_change}.
\begin{enumerate}
  \item Undeployed: The initial state of the instance lifecycle, the instance creation decision has been triggered by pre-warming or cold start. However, the instance does not yet physically exist in any worker node, though it can already receive invocations.
  \item Creating: The instance is in the process of creation. In this process, the image of the instance is downloaded from a remote repository and the runtime environment is initialized. The instance cannot execute any invocation until this process finish.
  \item Paused: The instance has been successfully created, or the invocations allocated to it have been completed, The instance is idle and waiting for new function invocations. The paused state is helpful for reducing cold starts since it is ready for immediately executing new invocations.
  \item Running: Once a function invocation is received, the instance transitions from the paused state to the running state. In this state, the instance is executing a function invocation.
  \item Killed: If the instance remains in the paused state beyond a specified keep-alive time, it is terminated, and its associated resources are released.
\end{enumerate}

\subsection{Function instance deployment}
In this section, newly created function instances are deployed after selecting available worker nodes (judged by Formula \eqref{eq:memory_limitation}) to minimize the network delay caused by data transmission.
In this paper, three strategies for function instance deployment are proposed.

\subsubsection{Dynamic load balancing deployment strategy (DLBDS)}
\begin{figure}[!h]
    \centering
    \includegraphics[scale = 0.6]{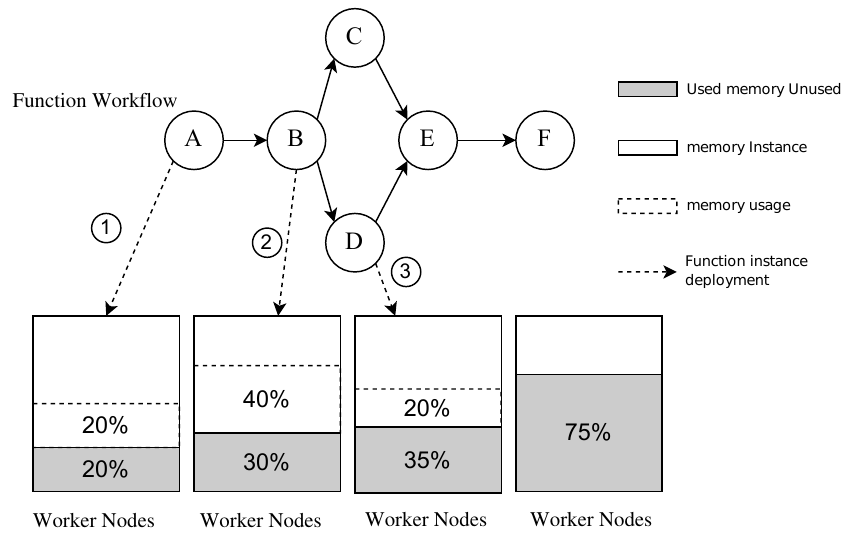}
    \topcaption{Dynamic load balancing deployment strategy diagram}
    \label{Fig:dynamicschedule}
\end{figure}

The DLBDS strategy evaluates the memory usage of all active worker nodes to deploy function instances.
It always selects the available worker node with the lowest memory usage to deploy the instance to be created.
This strategy avoids potential performance degradation due to imbalanced load, where a worker node may experience either excessive or insufficient load.
As the example shown in Figure \ref{Fig:dynamicschedule}, instances of functions A, B, and D are sequentially deployed to the worker nodes with the lowest memory usage according to the DLBDS strategy. After each instance is deployed, the memory usage of the worker nodes is updated for the next decision.
If some instances are released between the deployment of two function instances, the memory usage of the worker nodes is also updated.

\subsubsection{Affinity deployment strategy (ADS)}
The ADS strategy prioritizes deploying function instances with dependency relationships on the same worker node to enhance data locality, thereby reducing the network latency.
The locations of already deployed function instances are collected to support deployment decisions.
The ADS strategy is detailed in Algorithm \ref{alg:DRIAD_al}.
First, if a function is not invoked by any other functions, its instances can be deployed on any worker node with sufficient resources.
In this case, the worker node with the lowest memory usage is selected to ensure that subsequent function instances are more likely deployed to the same node.
Secondly, if a function may be invoked by others, the worker nodes where these functions are deployed are collected, and the first available node with sufficient resources is selected.

\begin{algorithm}[!ht]
    \begin{small}	
        \topcaption{Affinity Deployment Strategy (ADS)}
        \label{alg:DRIAD_al}
        \KwIn{$I_k$: Instance to be deployed, S: System state}
        \KwOut{$wn_w$: Selected worker node}
        \Begin{
            $f \gets$ The corresponding function of instance $I_k$\;
            $Pred \gets$ The predecessors of $f$\;
            $Aw \gets$ Get the available worker nodes for deploying instance $I_k$ from $S$\;
            \If{$Aw = \emptyset$} {
                $wn_w \gets$ Deploy $I_k$ on a new worker node\;
            }
            \Else{
                \If{$pred = \emptyset$} {
                    $wn_w \gets$ Deploy $I_k$ on the worker node with lowest memory usage\;
                }
                \Else{
                    $wn_w \gets$ Deploy $I_k$ on the first available worker node\;
                }
            }
            \Return{$wn_w$};
        }
    \end{small}
\end{algorithm}

\subsubsection{Free distribution deployment strategy (FDDS)}
The FDDS strategy is a function instance deployment method based on random deployment. It assigns instances to the first available worker nodes, without considering memory usage or network delay caused by communication. This strategy allows for faster selection of worker nodes, thereby creating instances earlier.

\subsection{Request routing}
The purpose of request routing is to assign each function invocation to a suitable instance according to the current system state, thereby avoiding cold starts and network latency.
Three function invocation allocation strategies are proposed.

\subsubsection{Shortest waiting time priority allocation strategy (SWPAS)}
This strategy aims to minimize the waiting time of functions (i.e., the gap between function invocation and function execution),
which is caused by function execution or unfinished instance creation.
Firstly, the SWPAS strategy collects all instances that can execute the function invocation.
Secondly, if there is an idle instance, the function invocation is directly allocated to this instance.
Otherwise, the function invocation is assigned to the instance with the shortest waiting time.
If the waiting time is larger than the cold start time, the invocation will be routed to a newly created instance.

\subsubsection{Shortest function elapsed time priority allocation strategy (SFEPAS)}

The basic idea of the SFEPAS strategy is to ensure that for each function invocation, there is at least one idle available instance before it is allocated.
The allocation of the function invocation is delayed until there exists an available instance.
Firstly, if no idle instance is available, a new instance is created immediately.
However, the invocation may not necessarily be allocated to this instance.
Secondly, once available instances are present, the invocation is allocated to the instance located on the worker node with the largest available memory.
Though this strategy may cause a cold start for the current function invocation, it may not increase the elapsed time.

\subsubsection{Minimum network communication priority allocation strategy (MNCPAS)}
The core idea of MNCPAS is to execute all function invocations in a workflow on a single worker node.
First, the MNCPAS strategy identifies the worker node with the maximum available memory and assigns the current function invocations to instances on that node.
Subsequently, if there are no current instances of successor functions available, it pre-warms instances on the same worker node.
If the memory of the current node is insufficient to create instances for all functions in the workflow, the node is bound to the workflow.
Once a function is finished, the corresponding instances are immediately released.
For subsequent function invocations within the workflow, the bound worker node is allocated directly.

\section{Experimental results}
This section involves conducting experiments to fine-tune parameters across various components of the algorithm, as well as evaluating the performance of the ICPS algorithm relative to current algorithms.
The Analysis of Variance (ANOVA) technique is adopted to analyze the results.
In the following, the experimental settings are introduced, followed by the results of parameter calibration and algorithm comparison.

\subsection{Experimental setting}

    \subsubsection{Experimental platform}

    The experiments are conducted on a platform equipped with an Intel(R) Xeon(R) Platinum 8260C CPU @ 2.30GHz, an NVIDIA GeForce RTX 4090 GPU, and 256GB of memory.
    The version of the operating system is Ubuntu 22.04.
    The algorithms are coded using Golang 1.22.3.
    Python 3.12.2 and PyTorch 2.2.2 are used for training the LSTM network.
    Communication between the models and algorithms is enabled using the HTTP protocol, with data transmitted in JSON format.

    We construct the serverless computing platform refer to \cite{kumari2024acpm}.
    This platform is written in Go language, including major functional modules such as system platform construction, configuration center, container module,  workflow instance creation module, and physical resource abstraction module.
    It provides instance management functions including managing the creation, execution, and termination of function instances, physical machine simulation for creating and allocating computing resources (limited to memory resources), configuring system parameters, and creating workflow instances.
    These modules and functionalities, simulate requests, scheduling, execution processes, and responses of workflow instances. The platform also simulates the allocation and release of computing resources. It simulates events such as the creation, execution, and termination of function instances that occur when the serverless computing platform serves workflows.
    In this platform, the relationships between functions are stored using a hash table, where the keys are function names and the values are custom data types used to store information such as function name, type, execution time, called function, calling function, etc.
    The platform adopts a unified external interface for the algorithm modules mentioned in this paper.
    Different methods for different algorithm modules are implemented based on the algorithm module interface, and then method selection is configured in the configuration center to achieve combinations of different methods across modules.

    The data of workflows in this experiment is sourced from the cold start tracing dataset of Alibaba Computing Service Center \cite{dataset}.
    The data information is stored in the backend database, including the name of the workflow, the dependency relationships among functions within the workflow, the arrival time of the workflow, names of functions, execution times of functions, and memory consumption of function invocations.
    Workflow requests are simulated by querying the backend database.
    The system simulates the concurrent invocation and execution of requests.
    Finally, the response times of workflows and resource utilization are generated.
    The information on function execution, instance deployment, cold starts, etc., is stored in the backend database.

    All algorithms in this experiment are implemented in the Go language. The combinations of system parameters and algorithm modules are configured via ini files. The platform generates experimental results under different conditions by reading various configuration files.

    \subsubsection{Experiment Parameter}
    The settings of experimental parameters are shown in Table \ref{fig:setting1}.

    In terms of generating workflows, two aspects are considered: concurrency level and depth.
    Workflows were derived from the dataset, characterized by arrival times within $[0, 800000)$ ms.
    The concurrency levels of workflows are set as $\{1, 5, 25, 50, 70\} \times 10^4$, representing the total number of workflows arriving within the specified time interval.
    The depths of the workflows are set as $\{1, 2, 5, 10, 20\}$.

    In terms of resource configuration, the number of worker nodes and the memory size of each worker node are considered.
    The number of worker nodes takes values from \{10, 20, 30\} and the memory size ranges in \{500, 1000, 3000\} MB.
    The two parameters are configured relative to the memory consumption of function invocations in the dataset.

    In terms of the network conditions, to simulate the network latency of a real-world serverless computing platform, four different levels of network delay \{2, 4, 6, 10, 15\} ms are set, relative to the execution time of functions in the dataset.

    For the LSTM network, the time step is set as 10 minutes, and the length of the time series is 36.
    The LSTM network just contains one hidden layer, and the number of neurons in this layer is 64.
    Adam is adopted as the model optimizer, and mean squared error (MSE) is employed as the loss function.
    When training the LSTM network, the learning rate is 0.001, the batch size is 32, and the number of epochs is 1000.

    \begin{table}[]
        \topcaption{Experimental parameter setting}
        \label{fig:setting1}
        \begin{center}
            \begin{tabular}{lll}
                \toprule 
                Parameter & Value & Unit \\ 
                \midrule 
                Concurrency level & $\{1, 5, 25, 50, 70\} \times 10^4$ & -\\
                Depth & $\{1, 2, 5, 10, 20\}$ & -\\
                Number of worker nodes & $\{10, 20, 30\}$ & -\\
                Memory size & $\{500, 1000, 3000\}$ & $MB$ \\
                Network delay & $\{2, 4, 6, 10, 15\}$ & $ms$ \\
                \bottomrule 
            \end{tabular}
        \end{center}
  \end{table}

    \subsubsection{Evaluation metrics}

    The ICPS algorithm aims to optimize the response time of workflows and the resource utilization. The objective is computed in Equation \ref{eq:targgget}.
    Relative Percentage Deviation (RPD) is used to evaluate the relative performance of algorithms.
    \begin{gather}
                RPD(\%) = \frac{\eta(\pi_B) - \eta(\pi_c)}{\eta(\pi_c)}\label{eq:rpdeta}
    \end{gather}
    where $\pi_c$ is the schedule generated by the currently conducted algorithm, and $\eta(\pi_c)$ is the objective value of this schedule.
    The optimal scheduling process is $\pi_B$, which can be estimated by the minimum of all compared algorithms.
    According to Formula \eqref{eq:rpdeta}, a small RPD value indicates that the current schedule is close to the optimal schedule.

\subsection{Parameter calibration}
    Three are three components in the ICPS algorithm that need to be calibrated: invocation concurrency prediction, function instance deployment, and request routing.
    Three strategies (FPCG, BPCG, CHSCG) are proposed for invocation concurrency prediction, and three strategies (DLBDS, ADS, FDDS) are designed for function instance deployment.
    In addition, three strategies (SWPAS, SFEPAS, MNCPAS) need to be calibrated in request routing.Therefore, there are 3$\times$3$\times$3=27 strategy combinations in total.
    For each strategy, the corresponding experiment is performed 10 times.
    Considering various settings of experimental parameters, there are 5$\times$5$\times$3$\times$3$\times$5$\times27\times10=303750$ tests conducted for parameter calibration.
    The best strategy combination is selected for constructing the ICPS algorithm.
    The experimental results are presented as follows.

    \begin{figure}[!h]
     \centering
     \subfigure[\text{\normalsize Comparison of invocation concurrency prediction strategies}]{
      \label{Fig:functionconcurrencygenerate3comp}
      \includegraphics[width=0.5\textwidth]{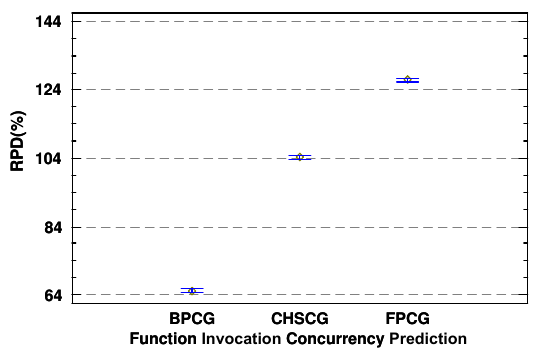}
     }
     \subfigure[\text{\normalsize Comparison of function instance deployment strategies}]{
      \label{Fig:functioninstancedeploy3com}
      \includegraphics[width=0.5\textwidth]{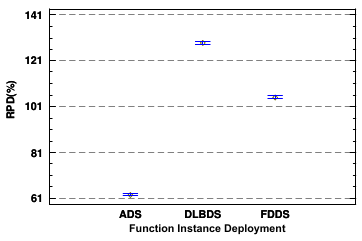}
     }
     \subfigure[\text{\normalsize Comparison of request routing strategies}]{
      \label{Fig:functioncallallocate3comp}
      \includegraphics[width=0.5\textwidth]{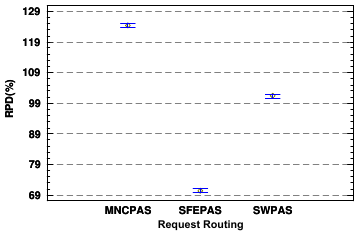}
     }
    \captionsetup{skip=5pt}
     \caption{Performance comparison of different strategies in different algorithm components with 95.0\% Tukey HSD confidence interval}
     \label{Fig:CC-2}
    \end{figure}

\subsubsection{\textbf{Instance calls concurrency prediction}}
Figure \ref{Fig:functionconcurrencygenerate3comp} shows the comparison of three strategies in invocation concurrency prediction with a 95\% Tukey HSD confidence interval.
The results indicate that the performance of the BPCG strategy is significantly better than the FPCG and CHSCG strategies since its RPD value is lowest.
The BPCG strategy first obtains prediction results of workflow concurrency and subsequently generates the function concurrency.
For workflow applications with branches, the BPCG can effectively avoid generating unnecessary function instances.
In contrast, the FPCG strategy generates instances for all functions in the workflow, which greatly increases resource consumption. Based on the above analysis, the BPCG strategy is chosen for predicting the concurrency of functions.

\subsubsection{\textbf{Function instance deployment}}
Figure \ref{Fig:functioninstancedeploy3com} shows the comparison of three strategies in function instance deployment with a 95\% Tukey HSD confidence interval.
It can be observed that the RPD of the ADS strategy is the lowest.
That means the ADS strategy performs better than the other two strategies.
In the ADS strategy, if a node with affinity has sufficient resources, deploying the function instance on that node can effectively avoid network latency caused by data transmission.
Therefore, the ADS strategy is adopted for function instance deployment.

\subsubsection{\textbf{Request routing}}
The comparison of strategies in request routing is shown in Figure \ref{Fig:functioncallallocate3comp}.
It indicates that the performance of the SFEPAS strategy is the best among the three strategies.
The SFEPAS strategy does not immediately allocate a function invocation. Instead, it attempts to match function invocations to appropriate idle instances.
It efficiently shortens the network latency while reducing the occurrence of cold starts. Therefore, the SFEPAS strategy is employed for request routing.

\vspace{0.1cm}

\subsection{Algorithm comparison}
To validate the effectiveness of the proposed algorithm in this paper, several existing algorithms are used for comparison, considering the similarity of the scenarios, algorithms, and objectives.
\begin{enumerate}[]
\item \textbf{EF-TTC \cite{MOAKHAR2024103890}}: The EF-TTC algorithm adopts the TTC mechanism, which is designed to address allocation issues and can generate effective solutions.
    The fundamental idea of the TTC algorithm is to construct a graph in which nodes represent users or servers and edges denote user preferences for servers and server priorities for users.
    The algorithm distinguishes between workflows submitted by different users, deploying function instances to servers using an affinity-based deployment approach to reduce network latency caused by data transmission. Regarding the cold start problem, the EF-TTC algorithm sets a fixed keep-alive time. If a new function instance needs to be started (cold start), the server must release sufficient resources (memory and cores).
    This is achieved by using the Least Recently Used (LRU) algorithm, which releases the recently least used function instance for deploying new function instances.
\item \textbf{PGP \cite{Performance-First}}: PGP is a prediction-based graph partitioning algorithm that aims to reduce the startup costs and execution latency of serverless workflows. In this algorithm, the functions of workflows are regarded as nodes in the graph, and the optimal location of function deployment is determined by the graph partitioning algorithm.
\item \textbf{CAS \cite{lifecycle-aware}}: Container lifecycle-aware scheduling strategies employ two key ideas to mitigate cold start issues: controlling request allocation based on different container lifecycle stages (start, run, pause), and determining when to create new containers or remove existing ones based on the current state of the containers. The algorithm has a similarity with the algorithm proposed in this paper in using container states as a guideline for function instance deployment.
\item \textbf{ACPM \cite{kumari2024acpm}}: The ACPM algorithm reduces cold starts by configuring containers at runtime. ACPM reduces cold starts through two stages: In the first stage, it uses an efficient LSTM to predict the number of containers that need to be pre-warmed in the future. In the second stage,  ACPM groups functions into single containers based on sandboxing, rather than creating containers for each function. The algorithm aims to reduce the frequency of cold starts,  balancing the response time and resource utilization.
\item \textbf{Keep-Alive \cite{9860368}}: Keep-Alive is a strategy adopted by serverless computing platforms like AWS to mitigate cold starts. This strategy involves keeping the containers active for a duration to reduce the need to launch a new container while processing a request. Thus, even if no requests arrive, Lambda function containers can remain active.
\item \textbf{Pool \cite{lin2019mitigating}}: Resource pooling is another strategy adopted by serverless computing platforms such as Fission and Knative. When a request arrives, the function can immediately be executed on one of these instances in the pool without waiting for a new container to start, thereby avoiding cold starts. Fission supports dynamically adjusting the size of the warm-up pool based on the amount of requests. When requests increase, Fission can automatically scale the warm-up pool to ensure enough instances are available to handle requests. In contrast, if requests decrease, Fission automatically scales the warm-up pool to save resources.
\end{enumerate}

The ICPS algorithm is compared to the above algorithms.
The settings of workflow concurrency, depth, number of worker nodes, memory size, and network delay are the same as in the previous section.
The experiments are conducted 10 times under each combination of parameters.
Therefore, there are $7\times$5$\times$5$\times$3$\times$3$\times$5$\times$10=78,750 tests conducted in algorithm comparison.


    \begin{figure}[!h]
     \centering
     \subfigure[Comparison in different concurrency levels]{
      \label{Fig:workflowconalg3}
      \includegraphics[width=0.49\textwidth]{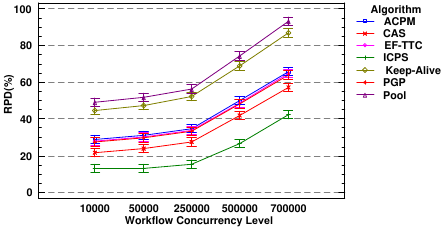}}
     \subfigure[Comparison in different depths of workflows]{
      \label{Fig:functioncallchaindepth3al}
      \includegraphics[width=0.49\textwidth]{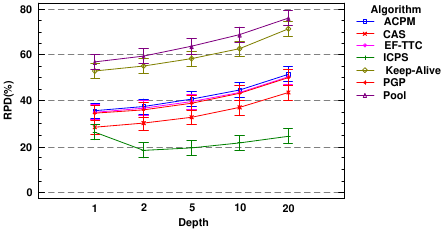}}
     \subfigure[Comparison in different resource configuration]{
      \label{Fig:workernumber3al}
      \includegraphics[width=0.49\textwidth]{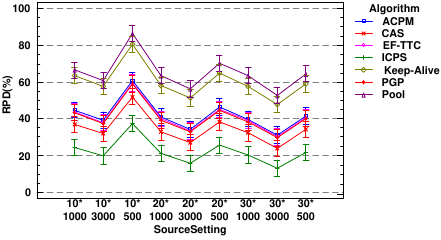}}
     \subfigure[Comparison in different network condition]{
      \label{Fig:networkdelay3al}
      \includegraphics[width=0.49\textwidth]{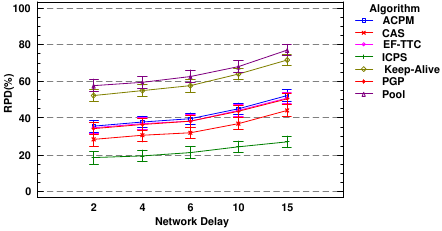}}
     \caption{Comparison of different algorithms in different conditions with 95.0\% Tukey HSD confidence interval}
     \label{Fig:DCITSresource}
    \end{figure}


\subsubsection{\textbf{Comparison in different concurrency levels}}
Figure \ref{Fig:workflowconalg3} illustrates the comparison of the ICPS algorithm and six other algorithms under varying levels of concurrency of workflows, with a 95\% Tukey HSD confidence interval.
The results indicate that the ICPS algorithm proposed in this paper outperforms the other algorithms across different levels of concurrency.
Compared to other comparison algorithms, the ICPS algorithm achieves the best performance by reducing cold starts through pre-warming and minimizing network latency through affinity-based deployment location selection.
The RPDs of all algorithms show an upward trend with the increasing workflow concurrency.
The reason is that as the concurrency of workflow increases, the available computing resources in the system decrease when functions are invoked. In some cases, function invocations are delayed until some instances transfer to the paused state from the running state.
However, if the concurrency of workflows does not exceed a certain threshold (i.e., 250000 in Figure \ref{Fig:workflowconalg3}), the upward trend in RPD values of all algorithms is gradual, with ICPS demonstrating the most moderate increase.
This is because the amount of resources can adequately meet the resource requirement of concurrent workflows at this stage.

\subsubsection{\textbf{Comparison in different depths of workflows}}
Figure \ref{Fig:functioncallchaindepth3al} shows the comparison between the ICPS algorithm and six other algorithms under different depths of workflows, with a 95\% Tukey HSD confidence interval.
In all cases, the ICPS algorithm outperforms other algorithms. The RPDs of the comparison algorithms show a significant upward trend as the depth increases because of the accumulation of cold start delays in the workflow.
However, the RPD of ICPS demonstrates a trend of initially decreasing, followed by an increase as the depth increases.
This is because, at a depth of 1 in the workflow (indicating that the workflow consists of only a single function), ICPS is unable to pre-warm function instances according to the dependency relationship.
When the workflow includes fewer functions, ICPS can accurately judge the branches and pre-warm proper numbers of function instances.
As the depth increases, misjudgments in branches by ICPS have a greater impact on response time, causing the RPD values to exhibit an upward trend.

\subsubsection{\textbf{Comparison in different resource configuration}}
The comparison between the ICPS algorithm and six other algorithms under different resource configurations is shown in Figure \ref{Fig:workernumber3al}. The results indicate that ICPS performs better than the other comparative algorithms regardless of the number of worker nodes and the memory size.

\subsubsection{\textbf{Comparison in different network condition}}
Figure \ref{Fig:networkdelay3al} shows the comparison between the ICPS algorithm and six other algorithms under different network conditions, with a 95\% Tukey HSD confidence interval.
The performance of ICPS is the best among all compared algorithms.
As the network delay increases, the RPD of ICPS shows a linear upward trend, whereas comparative algorithms exhibit a trend similar to ICPS under low network delay conditions, but a steeper upward trend under high network delay conditions. This indicates that ICPS performs better under high network delay conditions.

\section{Conclusions and Future Work}
In this paper, a response time minimization problem with minimal resource waste is studied for dynamic workflows with branches in serverless computing.
The ICPS algorithm is proposed with four components.
It configures the numbers of function instances by invocation concurrency prediction to be pre-warmed and realizes affinity-based instance deployment.
The results illustrate the effectiveness of the ICPS algorithm in various cases.
Especially, ICPS performs better in high network delay conditions.

There are still some topics that need further research.
For example, when different function instances are co-located on the same worker node, interference between them can degrade performance.
Optimizing the response time of workflows in such scenarios presents a significant challenge.
\bibliographystyle{IEEEtran}
\bibliography{reference}


\begin{IEEEbiography}
[{\includegraphics[width=1in,height=1.25in,clip,keepaspectratio]{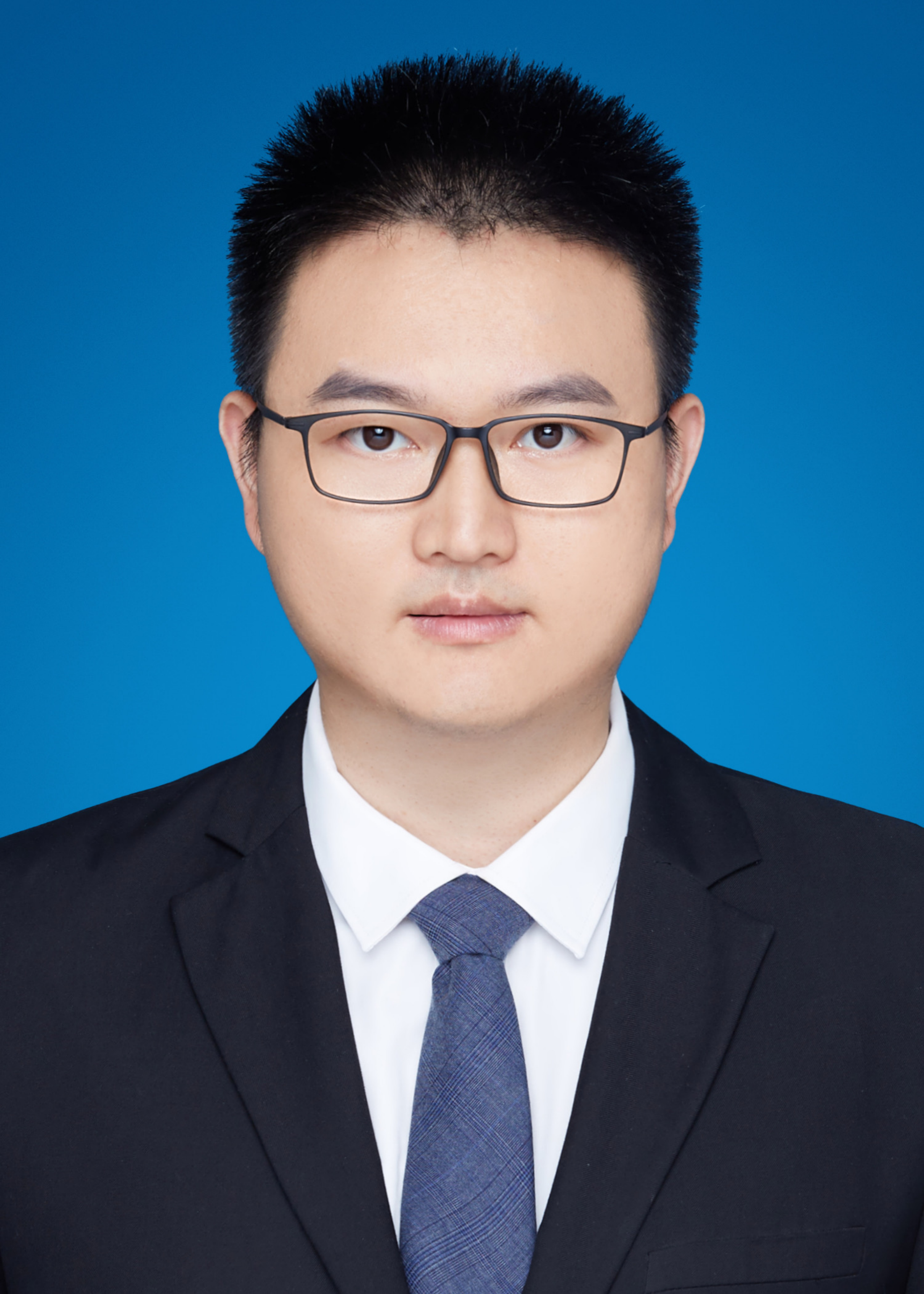}}]{Long Chen}
received his B.Sc. and Ph.D. degrees in Computer Science and Engineering from Southeast University, Nanjing, China, in 2009 and 2018, respectively.
He is currently serving at the School of Computer Science and Engineering at Southeast University, Nanjing, China.
He has published more than 30 papers in international journals and conferences, such as \emph{IEEE Transactions on Services Computing}, \emph{IEEE Transactions on Cloud Computing}, \emph{IEEE Transactions on Automation Science and Engineering}.
His main research interests include Cloud Computing, Edge Computing, and Service-oriented Computing. 
\end{IEEEbiography}

\begin{IEEEbiography}
[{\includegraphics[width=1in,height=1.25in,clip,keepaspectratio]{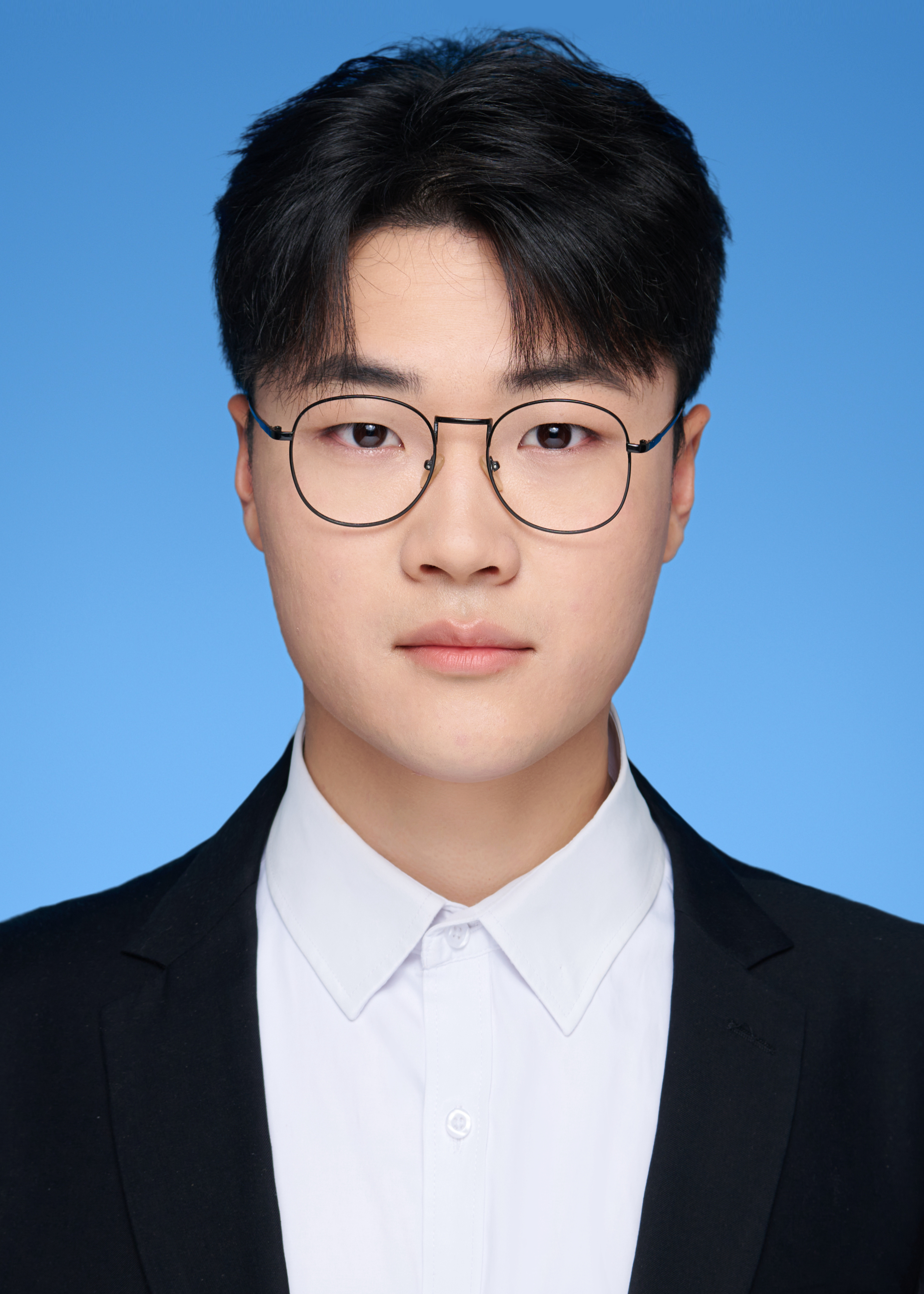}}]{Xinshuai Hua}
is currently pursuing the M.S. degree in the School of Computer Science and Engineering at Southeast University, Nanjing, China.
His main research interests include Edge Computing and UAV wireless communications.\vspace{-10mm}
\end{IEEEbiography}

\begin{IEEEbiography}
[{\includegraphics[width=1in,height=1.25in,clip,keepaspectratio]{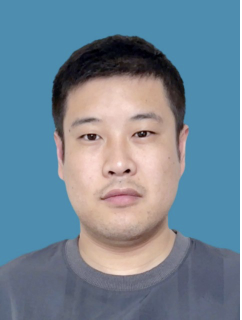}}]{Wenshuai Li}
received his B.Sc. degree in Computer Science and Engineering from Southeast University, Nanjing, China, in 2024.
His main research interests include Cloud Computing and Serverless Computing.\vspace{-10mm}
\end{IEEEbiography}

\begin{IEEEbiography}[{\includegraphics[width=1in,height=1.25in,clip,keepaspectratio]{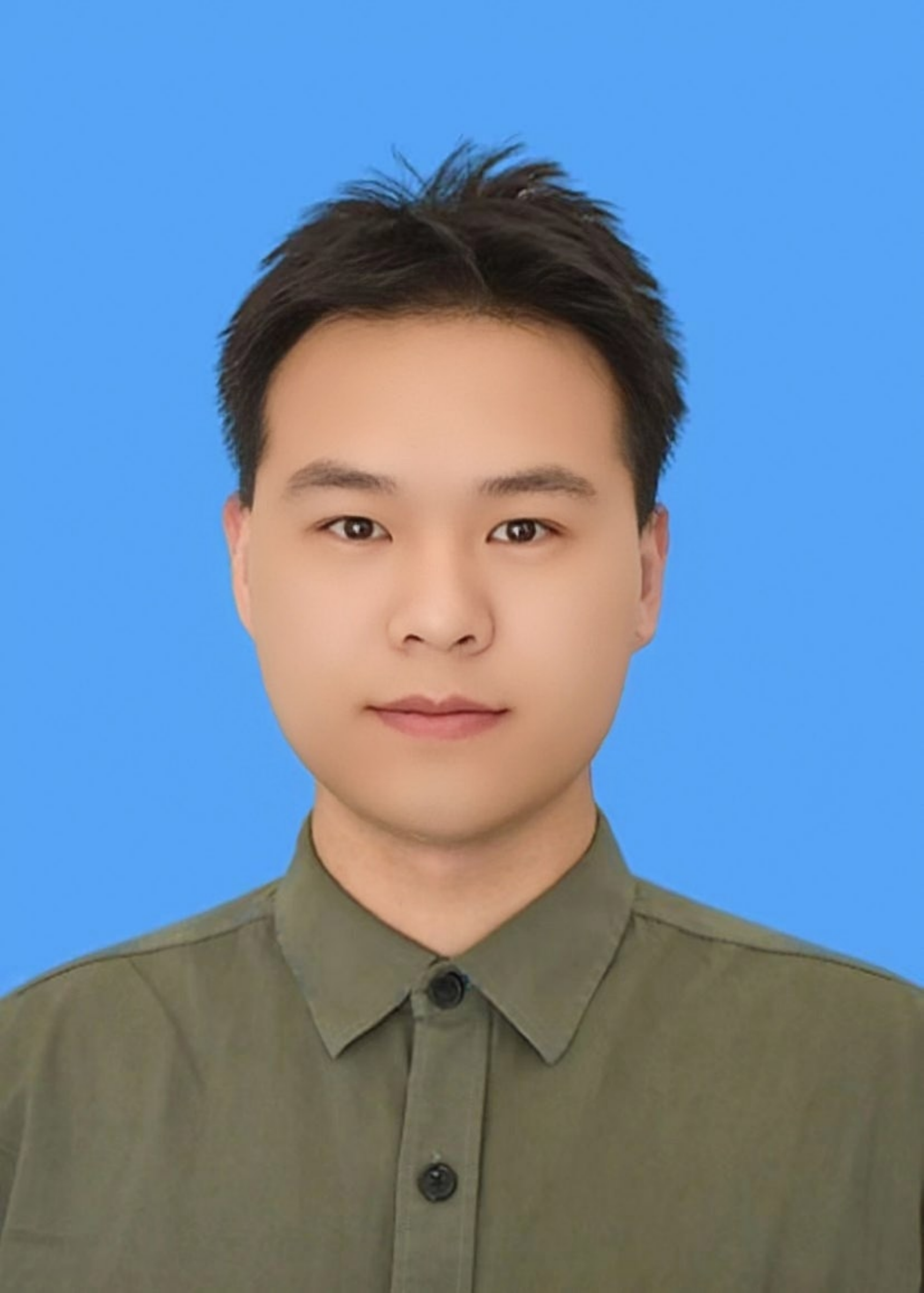}}]{Jinquan Zhang}
received his Ph.D. degree in the College of Computer Science and Engineering from Southeast University, Nanjing, China, in 2024.
He is currently a lecturer in the School of Computer Science and Technology, Guangdong University of Technology, Guangzhou, China.
He is the author or co-author of about 10 papers in international journals and conferences, such as \emph{IEEE Transactions on Services Computing}, \emph{IEEE Transactions on Cloud Computing}, \emph{Future Generation Computer Systems}.
His research interests focus on cloud resource management and serverless computing.
\end{IEEEbiography}

\begin{IEEEbiography}
[{\includegraphics[width=1in,height=1.25in,clip,keepaspectratio]{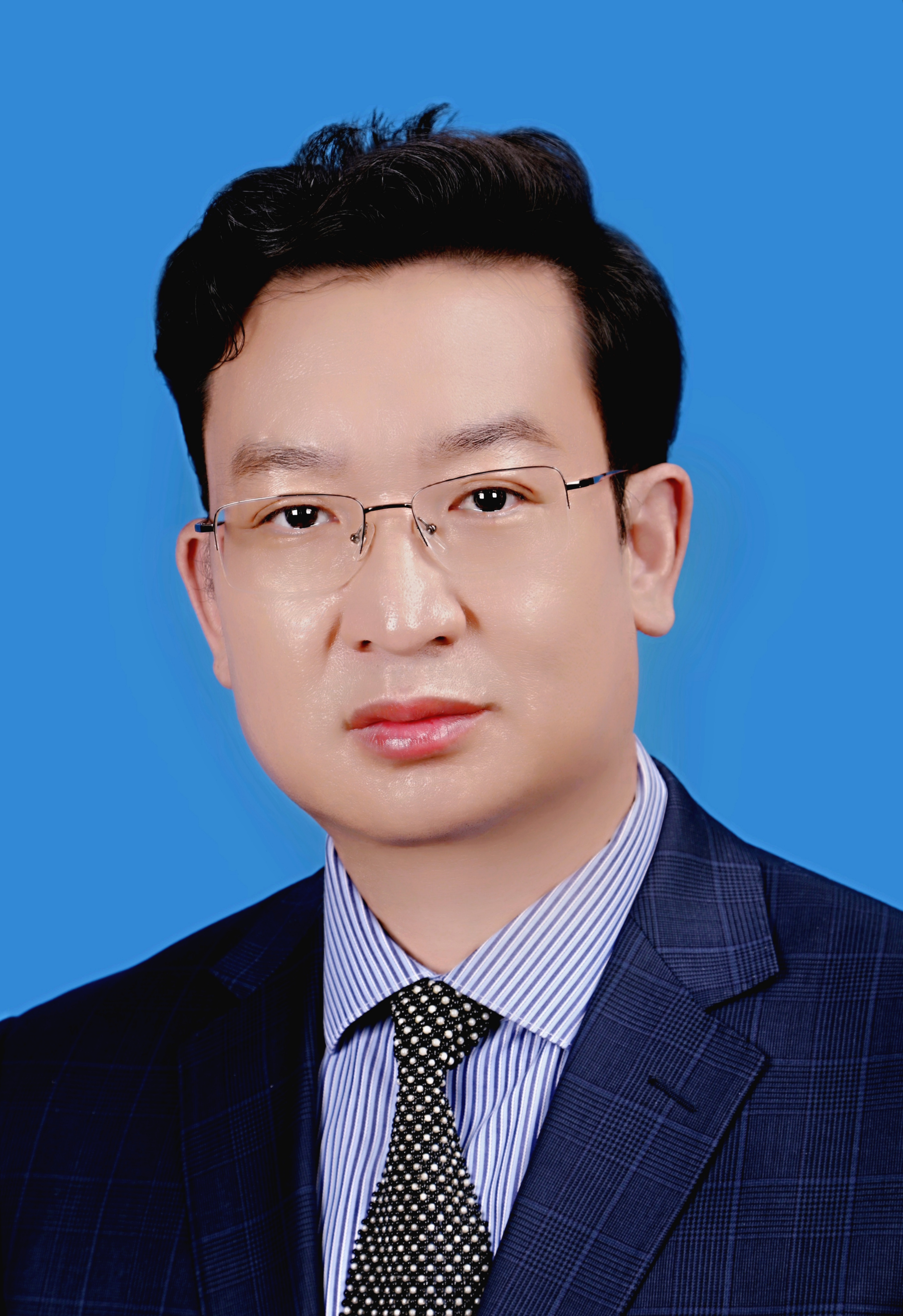}}]{Xiaoping Li}
    (M’09-SM’12) received his B.Sc. and M.Sc. degrees in Applied Computer Science from Harbin University of Science and Technology in 1993 and 1999, respectively, and his Ph.D. degree in Applied Computer Science from Harbin Institute of Technology in 2002.
    He was a distinguished professor with the School of Computer Science and Engineering, Southeast University, Nanjing, China. 
    He is currently a full professor and the head of the School of Computer Science and Technology, Guangdong University of Technology, Guangzhou, China.
    He is the author or coauthor of more than 100 academic papers, some of which have been published in international journals such as \emph{IEEE Transactions on Computers, IEEE Transactions on Parallel and Distributed Systems, IEEE Transactions on Services Computing, IEEE Transactions on Cybernetics, IEEE Transactions on Automation Science and Engineering, IEEE Transactions on Cloud Computing, IEEE Transactions on Systems, Man and Cybernetics: Systems, Information Sciences, Omega, European Journal of Operational Research, International Journal of Production Research, Expert Systems with Applications, and Journal of Network and Computer Applications}. 
    His research interests include Scheduling in Cloud Computing, Scheduling in Cloud Manufacturing, Service Computing, Big Data, and Machine Learning.
\end{IEEEbiography}

\begin{IEEEbiography}[{\includegraphics[width=1in,height=1.25in,clip,keepaspectratio]{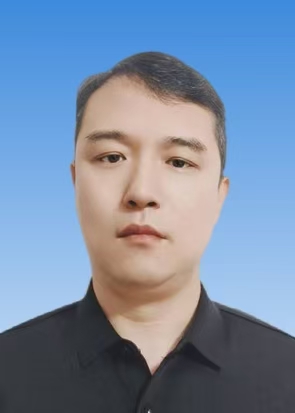}}]
{Shijie Guo} is a senior engineer in the College of Electronic Engineering, National University of Defense Technology, Hefei, China. 
He received his M.E degree from the Electronic Engineering Institute in 2012 and has published more than 30 papers. 
His research interests include intelligent software testing and digital testing simulation technology.
\end{IEEEbiography}

\vfill

\end{document}